\documentclass[preprint,showpacs,amsmath,amssymb,superscriptaddress,nofootinbib]{revtex4}

\usepackage{graphicx}
\usepackage{epsfig}
\usepackage{dcolumn}
\usepackage{bm}

\newcommand{\beq}{\begin{equation}}
\newcommand{\eeq}{\end{equation}}
\newcommand{\bea}{\begin{eqnarray}}
\newcommand{\eea}{\end{eqnarray}}

\begin{document}

\title {Rotational damping in a multi-$j$ shell particles-rotor model}

\author{Lu Guo}
\email{guolu@mcs.ibaraki.ac.jp} \affiliation {Department of
Mathematical Science, Ibaraki University, Mito 310-8512, Ibaraki,
Japan}
\author{Jie Meng}
\affiliation{School of Physics, Peking University, Beijing 100871,
China}
\author{En-Guang Zhao}
\affiliation {Institute of Theoretical Physics, Chinese Academy of
Sciences, Beijing 100080, China}
\author{Fumihiko Sakata}
\affiliation {Institute of Applied Beam Science, Graduate School
of Science and Engineering, Ibaraki University, Mito 310-8512,
Ibaraki, Japan}
\date{\today}
\begin{abstract}
The damping of collective rotational motion is investigated by
means of particles-rotor model in which the angular momentum
coupling is treated exactly and the valence nucleons are in a multi-$j$ shell
mean-field. It is found that the onset
energy of rotational damping is around 1.1 MeV above yrast line, and
the number of states which form rotational band structure is thus
limited. The number of calculated rotational bands around 30 at a
given angular momentum agrees qualitatively with experimental
data. The onset of rotational damping takes place gradually as a
function of excitation energy. It is shown that the pairing
correlation between valence nucleons has a significant effect on
the appearance of rotational damping.
\end{abstract}
\maketitle

\section {introduction}
The experimentally observed rotational bands often lie in the
region near yrast line and are described as particle-hole
excitations in the mean-field. At higher excitation energy above
the yrast line, it does not necessarily form rotational band
structure due to the damping of collective rotational motion
\cite{GA82,BL86}. When the rotational damping takes place, E2
transition from an excited state spreads out over many final
states. The gamma-rays which are emitted from the above excited
region can not be distinguished as discrete peaks, thus forming a
quasi-continuum spectra. Experimentally rotational damping has
been studied through the analysis of quasi-continuum spectra
\cite{JC85,JE86,FS86,FS87,FS88}. Recent experimental progress in
high precision three-dimensional gamma-ray correlation
measurements makes it possible to study various features of
collective rotational motion in the regions of discrete rotational
bands and damped rotational motion
\cite{BH92,TD96,BH921,BH922,BH93,SL95,SF99,FSS02}. In particular,
the newly developed fluctuation analysis method \cite{BH92,TD96}
has presented the number of rotational bands existing in a
rare-earth nucleus is only about 30 at a given angular momentum,
thus confirming the occurrence of rotation damping.

Early theoretical studies on rotational damping
\cite{GA82,BL86,TG89,JL91} showed that with the increase of level
density the off-diagonal residual interaction becomes effective to
cause mixing of many-particle many-hole configurations in the
rotating mean-field. Since different configurations respond
differently to the Coriolis force, the configuration mixing
results in a dispersion of rotational frequency within each energy
eigenstate, implying a damping of the collective rotational
motion. However, in these works \cite{GA82,BL86,TG89,JL91}, the
assumption that configuration mixing is described by the general
statistical theory of random matrices has been used to treat the
E2 strength function associated with the damped rotational motion.

Recent studies on the microscopic structure and mechanism of
rotational damping have been done extensively for normally
deformed and superdeformed nuclei in the cranked Nilsson
mean-field combined two-body residual interaction
\cite{MM93,MM971,MM972,KY97,KY98,MM99,TD01,KY01}. In these
discussions, the excited rotational bands are described as
intrinsic many-particle many-hole excitations in the cranking
Nilsson mean-field. In order to obtain rotational damping, the
bands are mixed by two-body residual interaction. Furthermore, the
cranking model is semi-classical and the angular momentum is not a
good quantum number.

One can also use the particles-rotor model (PRM) to
obtain the band structure as well as rotational damping and to
study the coupling between single-particle degree of freedom and
collective rotational motion. Moreover as noted in Ref.
\cite{MM971}, PRM is more appropriate on angular momentum coupling
in wave functions and E2 transition matrix elements, since angular
momentum is not a conserved quantity in cranking model. With many
particles in a single-$j$ shell, a systematic comparison of cranking
model and PRM has been done and the pair-correlation transition in
rotating nuclei has been investigated in Ref.~\cite{meng93}. In
Ref.~\cite{FM96}, PRM has been used to examine the quality of
the tilted axis cranking and its interpretation. In particular, by
correctly treating the angular momentum coupling in triaxial PRM,
a new phenomenon-chiral doublet bands has been
predicted~\cite{FM97} and verified later by experiment.
Therefore it will be interesting to study the rotational damping
by PRM. The nuclear chaotic behavior and its connection
with rotational damping have been
addressed in cranking model and PRM with single-$j$ shell \cite{AT95, SA90, YS04}. The purpose
of present work is to study the rotational damping by PRM, in
which the angular momentum is a good quantum number and the pairing
correlation will be included explicitly. The mechanism of
rotational damping and E2 transition property from the region
of discrete rotational band into the regime of damped rotational
motion will be investigated in a multi-$j$ shell PRM.

\section {Formulation}

A nucleus can be visualized as a rotor coupled with a few valence
nucleons, which move more or less independently in the deformed
potential of core composed by the rest of the nucleons, i.e., the
particles-rotor model (PRM). The Hamiltonian of PRM is expressed
as the sum of an intrinsic part and a collective part \beq
 H_{\textrm {PRM}}=H_{\textrm {intr}}+H_{\textrm {coll}},
\eeq where $H_{\textrm {intr}}$ describes microscopically the
motion of valence particles near Fermi level and $H_{\textrm
{coll}}$ is the collective rotation of core. The intrinsic
Hamiltonian is taken as
\beq H_{\textrm
    {intr}}=\sum_{j_1m_1j_2m_2}\langle
     j_1m_1|-8\kappa\sqrt{\pi/5}Y_{20}|j_2m_2\rangle
     a_{j_1m_1}^{\dagger}a_{j_2m_2} - G\sum_{j^\prime m^\prime jm}
     a_{j^\prime m^\prime} ^\dagger a_{j^\prime \bar m^\prime} ^\dagger
     a_{j\bar m} a_{jm},
\eeq
 where $a_{jm}^\dagger$ and $a_{jm}$ are the one-particle
creation and annihilation operators and a residual pairing
interaction has been included. In order to describe many-particle
many-hole excitations associated with rotational damping, we
include some different-$j$ shells in the deformed mean-field of
valence nucleons. The single-particle energy in the deformed
mean-field is written as \cite{SP1,SP2,AT95}
 \beq
    \sum_{j_1m_1j_2m_2}\langle
      j_1m_1|-8\kappa\sqrt{\pi/5}Y_{20}|j_2m_2\rangle
      a_{j_1m_1}^{\dagger}a_{j_2m_2} =\sum_j\Bigl\{ R_j+
      \sum_{m}\kappa\frac{3m^2-j(j+1)}{j(j+1)}a_{jm}^{\dagger}a_{jm}\Bigr\},
\label{sp}
\eeq
 where $R_j$ is a parameter to indicate the relative energy between
different-$j$ shells. The deformation parameter $\kappa$ is
related to the quadrupole deformation $\beta$ through \cite{YS04,
JAS98} \beq \kappa \simeq 0.16\hbar\omega_0(N+3/2)\beta, \eeq
where $\hbar\omega_0$ is harmonic oscillator frequency of the
deformed potential and $N$ the quantum number of the major shell.
For example, in the case of $i_{13/2}$ shell, $k=2.5$ MeV
approximately corresponds to $\beta\sim$ 0.3, and $h_{9/2}$ shell
corresponds to $k=2.2$ MeV. The two-body correlation between
valence particles is taken as pairing interaction with strength
parameter $G$. Since we solve two-body interaction exactly, it
will contribute to both particle-particle channel and
particle-hole channel.

The spin $\vec I$ of nucleus is the sum of angular momentum $\vec
R$ of core and $\vec J$, the sum of angular momentum of valence
nucleons. The collective Hamiltonian can be expressed as
 \beq
  H_{\textrm {coll}}=\frac{R_1^2+R_2^2}{2{\cal J}}
  =\frac{I^2-I_3^2}{2{\cal J}}+\frac{J^2-J_3^2}{2{\cal J}}
  -\frac{I_+J_-+I_-J_+}{2\cal J},
 \eeq
where $H_{\textrm {rot}}=(I^2-I_3^2)/2{\cal J}$ describes the
nuclear collective rotation. The recoil term $H_{\textrm
{rec}}=(J^2-J_3^2)/2{\cal J}$ acts only on valence nucleons and
contains one-body and two-body terms if there is more than one
particle outside the core. Coriolis interaction $H_{\textrm
{cor}}=-(I_+J_-+I_-J_+)/2\cal J$ couples the collective rotation
and the single particle motion.

The eigenfunctions of particles-rotor Hamiltonian can be expanded as
 \beq
\Psi_{IM}^\alpha=\sum_{K}C_K^{\alpha}\varphi_{IMK}^{\alpha}
 \eeq
with the mixing coefficient $C_K^{\alpha}$, where the symmetrized
basis is given by
 \beq
\varphi_{IMK}^{\alpha}=\sqrt{\frac{2I+1}{16\pi^2(1+\delta_{K0})}}\bigg\{ {\cal D}^{I
\ast}_{MK}(\Omega)\phi_{K\alpha}^{(1,2,\cdots N)}+
 (-)^{I+K}{\cal D}^{I \ast}_{M-K}(\Omega)\phi_{\bar K \alpha}^{(1,2,\cdots N)}\bigg\}.
 \eeq
${\cal D}^{I}_{MK}(\Omega)$ is the usual rotation matrix and
$\phi_{K\alpha}^{(1,2,\cdots N)}$ the $N$-body anti-symmetrized
wave function of valence particles.

The stretched E2 transition probability from an initial nuclear
state $\alpha$ at spin $I+2$ to a final state $\alpha^\prime$ at
$I$ is
 \beq B(E2,\alpha I+2\rightarrow \alpha^\prime I)
 =  \frac{5}{16\pi}Q_0 ^2 M^2_{\alpha I+2, \alpha ^\prime I},
 \eeq
where the amplitude
\beq M_{\alpha I+2, \alpha ^\prime
I}=\sum_{KK^\prime} C_K^\alpha(I+2)
C_{K^\prime}^{\alpha^\prime}(I) \langle IK^\prime 20|I+2
K\rangle\delta_{KK^\prime}.
 \eeq

We neglect the minor contribution of non-collective $E_2$
transition from valence particles since the electric quadrupole
moment $Q_0$ is much larger than that from valence particles. The
$CG$ coefficient $\langle IK20|I+2 K\rangle$ represents the
coupling of angular momenta in the intrinsic frame. The normalized
E2 transition probability $ S_{\alpha I+2, \alpha ^\prime I}$ is
defined as \beq S_{\alpha I+2, \alpha ^\prime I}\equiv M ^2
_{\alpha I+2, \alpha ^\prime I}, \eeq which satisfies $\sum
_{\alpha ^\prime}S_{\alpha I+2, \alpha ^\prime I}=1$ from an
initial state $\alpha$ to many final states $\alpha^\prime$.

\section{Results and discussions}

In present PRM, we consider eight valence particles in a multi-$j$
shell coupled with a deformed core. The calculation has been performed
with the parameters: the pairing strength $G=0.45$
MeV, the moment of inertia $\cal J$=76.0 $\hbar ^2{\textrm
MeV}^{-1}$, $\kappa=2.5$ MeV for $i_{13/2}$ intruder orbit and
$\kappa=2.2$ MeV for other shells. These parameters have been used in earlier studies
and are considered to be reasonable \cite{MM971, AT95, YS04}.

In order to imitate the level scheme of Nilsson diagram
\cite{PR80}, we include the shells $h_{9/2}$, $i_{13/2}$,
$p_{3/2}$, $f_{5/2}$ and $p_{1/2}$ in the deformed mean-field. The
relative energies $R_j$ between different-$j$ shells are
determined according to spherical Nilsson level diagram.
Table~\ref{tab1} gives the relative energy $R_j$ between shell
$h_{9/2}$ and other shells $i_{13/2}$, $p_{3/2}$, $f_{5/2}$ and
$p_{1/2}$, e.g. it has been taken as 0.1 $\hbar\omega_0$ between
shells $h_{9/2}$ and $i_{13/2}$. The choice of these parameters
for the present single-particle levels is similar as the neutron
levels in nucleus $^{168}$Yb. Of course, the valence proton excitations
should be included as well to fully understand nucleus $^{168}$Yb.
However, if both proton and neutron excitations are taken into
account, the Hamiltonian matrix will be too big to be diagonalized
exactly. In Refs.~\cite{FM96,FM97}, only one proton and
one neutron are coupled with the rotor, and the many-particle
and many-hole excitations are thus missing.

\begin{figure}
\epsfxsize=12.0cm \centerline{\epsffile{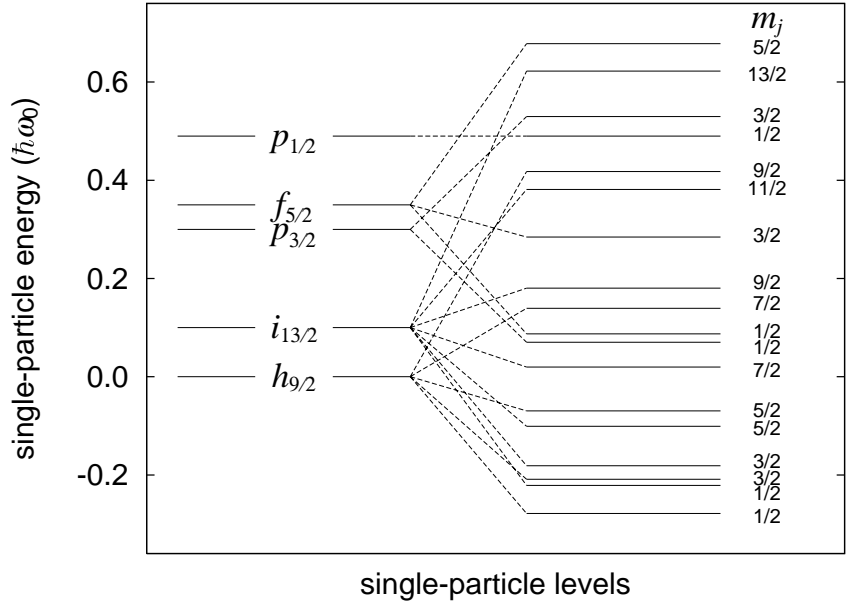}}
\caption{Single-particle level scheme in the mean-field of valence
particles. The spherical shells on the left split up to the
structure on the right due to the deformation with the third
component $m_j$ of angular momentum. }
\label{ref1}
\end{figure}

The level scheme thus obtained is shown in Fig. \ref{ref1}, in
which the spherical shells and their splitting due to deformation
are presented. On the right the third component $m_j$ of angular
momentum is given.

After defining the single-particle basis, the many-particle
many-hole excitations within a given excitation energy form the
shell model basis states. The configuration space is spanned by
various excitations of eight valence particles in the deformed
mean-field. The off-diagonal Hamiltonian from pairing correlation,
recoil term and coriolis interaction will mix states not only in
the same-$j$ shell, but also in different-$j$ shells
\cite{Guo1,Guo2}. They are responsible for the many-particle
many-hole excitations. In the model diagonalization for each spin
$I = 0-60$ $\hbar$, the configuration truncation is done with the
excitation energy $\sim 0.6$ $\hbar\omega_0$, which corresponds to
the lowest 2000 many-particle many-hole configurations. Most of
these configurations has $1p-1h$ to $4p-4h$ characters and few of
them is with $5p-5h$ to $6p-6h$ characters.

\begin{figure}
\epsfxsize=12.0cm \centerline{\epsffile{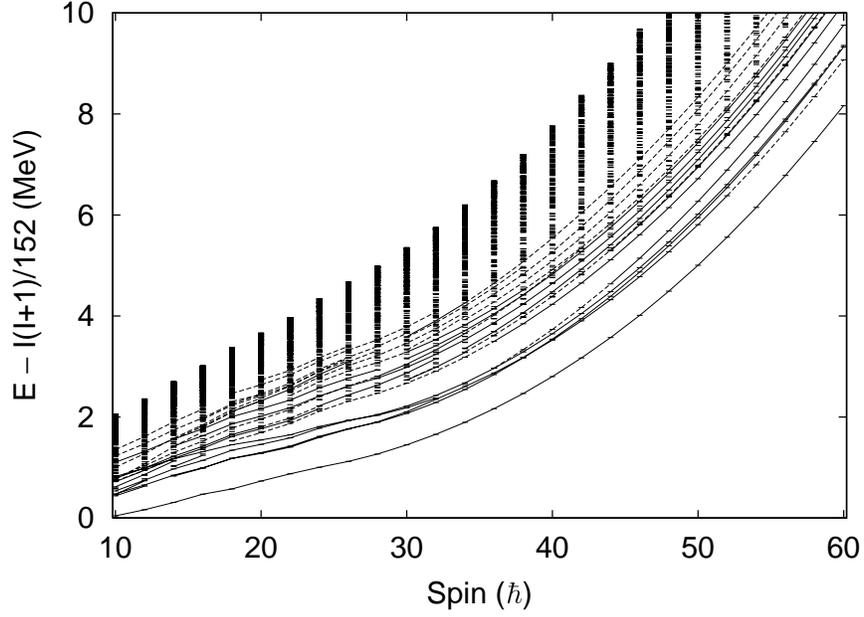}} \caption{The
calculated nuclear states are plotted with small horizontal bars.
A reference energy $I(I+1)/2{\cal J}$ with $\cal J$=76.0 $\hbar
^2{\textrm MeV}^{-1}$ is subtracted. Solid lines connecting the
energy levels represent the strong E2 transitions which have the
normalized strength $S_{\alpha I+2, \alpha^\prime I}$ larger than
0.707. Dashed lines are the weaker E2 transitions with normalized
strength between 0.5 and 0.707.} \label{energy}
\end{figure}

Fig.~\ref{energy} displays the calculated nuclear states with
small horizontal bars. Strong E2 transitions satisfying condition
$S_{\alpha I+2,\alpha ^\prime I}>1/\sqrt{2}=0.707$ are plotted
with solid lines. Weaker transitions defined by condition
$0.5<S_{\alpha I+2,\alpha ^\prime I}<0.707$ are presented with
dashed lines. One may observe that most of the strong E2
transitions lies in the region near yrast line, where the
rotational band structure is identified. At higher excitation
energy the transitions become much weaker and E2 transition from
an initial state may spread out over many final states, implying
the disappearance of rotational band structure.

\begin{figure}
\epsfxsize=8.0cm \centerline{\epsffile{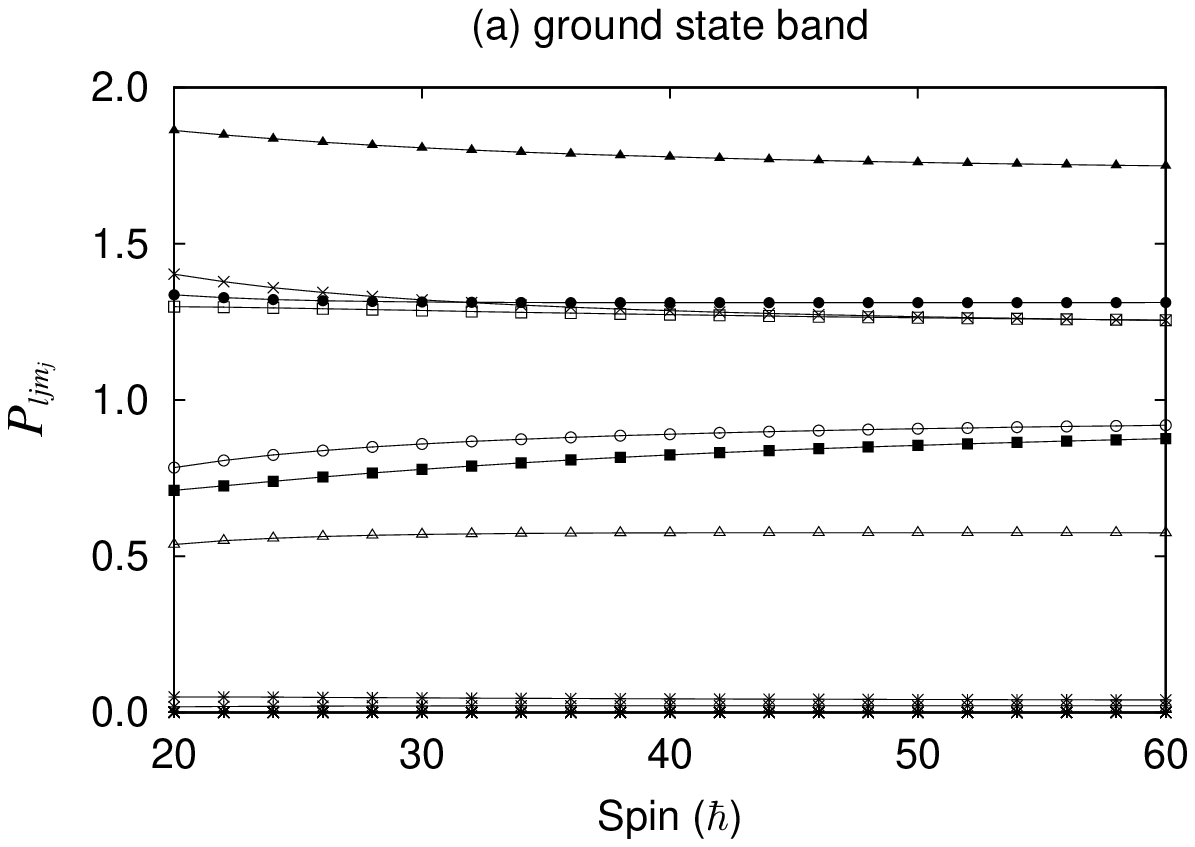}}
\centerline{\epsffile{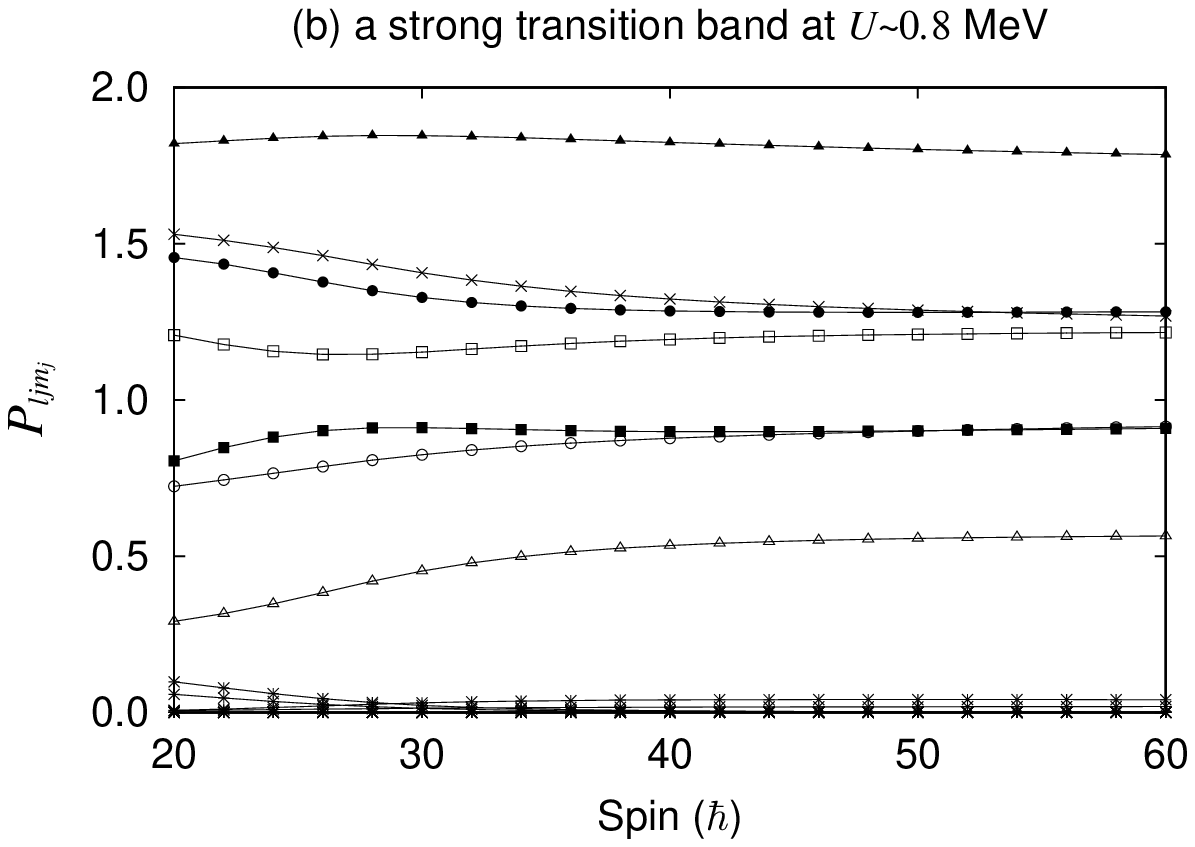}}
\centerline{\epsffile{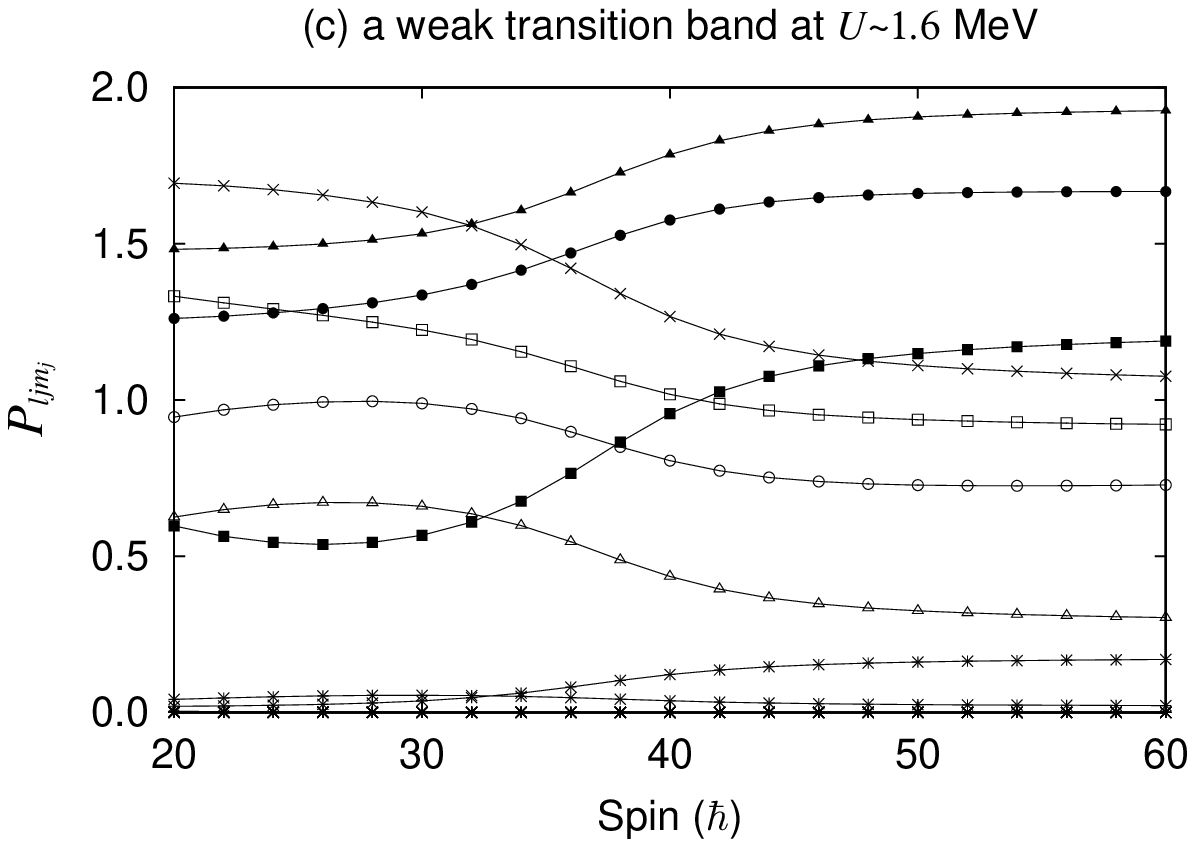}} \caption{Occupation number of
particle for various single-particle basis $[ljm_j]$ as a function
of spin for (a) ground state band; (b) a strong transition band at
excitation energy $U \sim 0.8$ MeV; (c) a weak transition band at
excitation energy $U \sim 1.6$ MeV. The lines with filled upper
triangle, cross, filled circle, open square, open circle, filled
squire and open upper triangle represent the single-particle basis
states $[5\frac{9}{2}\frac{1}{2}]$, $[6\frac{13}{2}\frac{1}{2}]$,
$[5\frac{9}{2}\frac{3}{2}]$, $[6\frac{13}{2}\frac{3}{2}]$,
$[6\frac{13}{2}\frac{5}{2}]$, $[5\frac{9}{2}\frac{5}{2}]$ and
$[6\frac{13}{2}\frac{7}{2}]$ separately. The lines with star
denote the occupancy on other single particle basis. }
\label{Occpro1}
\end{figure}

In order to understand the nature of nuclear state, a quantity,
occupation number of single-particle basis $[ljm_j]$
in nuclear state $\alpha$, is defined as
 \beq
    P_{ljm_j} = \sum_{\alpha(ljm_j)} {C_K^\alpha}^2,
 \eeq
  satisfying condition that the sum
of various occupancy on single-particle basis is equal to the
number of valence particles
 \beq
  \sum_{ljm_j}P_{ljm_j}=N .
 \eeq
Here we use quantum numbers $[ljm_j]$ to denote the single
particle basis states in Fig.~\ref{ref1}. Fig.~\ref{Occpro1} gives
the occupation number of particle for various single-particle
basis $[ljm_j]$ as a function of angular momentum for (a) ground
state band; (b) a strong transition band at excitation energy $U
\sim 0.8$ MeV; (c) a weak transition band at excitation energy $U
\sim 1.6$ MeV. The lines with filled upper triangle, cross, filled
circle, open square, open circle, filled squire and open upper
triangle represent the single-particle basis states
$[5\frac{9}{2}\frac{1}{2}]$, $[6\frac{13}{2}\frac{1}{2}]$,
$[5\frac{9}{2}\frac{3}{2}]$, $[6\frac{13}{2}\frac{3}{2}]$,
$[6\frac{13}{2}\frac{5}{2}]$, $[5\frac{9}{2}\frac{5}{2}]$ and
$[6\frac{13}{2}\frac{7}{2}]$ separately. Since the occupancy on
other higher single particle basis is not important for the low
excited states, we use the lines with star to represent the
occupancy on those single particle basis. One may observe for the
bands satisfying strong transition condition as shown in
Fig.~\ref{Occpro1} (a) and (b), the occupation number of particle
is nearly independent of angular momentum. With the increase of
excitation energy, the dependence of occupancy on spin becomes
more sensitive, as shown for the band with weak transition in
Fig.~\ref{Occpro1} (c). Since the occupancy characterizes the
wave function property of nuclear state, these figures clearly
indicate how the rotational band structure starts to disappear
gradually due to the change of wave function in nuclear states.

In the experiments, strong E2 transitions are observed as discrete
peaks in the gamma-ray spectra and the rest of transitions shows
up as quasi-continuum spectra which contain transitions summed
over many final states. For such a situation, it is useful to
represent the E2 transition property by means of strength
distribution function. The strength function for a state $\alpha$
at spin $I+2$ is given by \cite{MM971}
 \beq
    S_{\alpha,I+2}(E_{\gamma})=\sum_{\alpha^\prime}S_{\alpha I+2,
     \alpha^\prime I}\delta(E_\gamma -E_{\alpha
     I+2}+E_{\alpha^\prime I}).
 \eeq
 The fragmentation of E2 strength function is the
rotational damping phenomenon. In order to quantify the onset of
rotational damping, the branch number is defined as
 \beq
    n_{\textrm{branch}} (\alpha I+2) \equiv
     (\sum_{\alpha^\prime}S_{\alpha I+2, \alpha^\prime I}^2)^{-1}.
 \eeq
It counts effectively the branch number of E2 transitions from an
initial state $\alpha$ at $I+2$ to the final states
$\alpha^\prime$, which are allowed by the selection rules of the
gamma-ray radiation. For a case where a given state decays to only
one final state at the lower spin, $n_{\textrm {branch}}$ is equal
to one. If there are two possible transitions with equal
probability, the branch number is equal to two. In other words,
$n_{\textrm {branch}}<2$ implies that the transition from a given
state mainly decays to one final state, where the discrete
rotational band structure is identified. In contrast, $n_{\textrm
{branch}}>2$ states that the transitions spread over two or more
final states, where the rotational damping takes place.

\begin{figure}
\epsfxsize=12.0cm \centerline{\epsffile{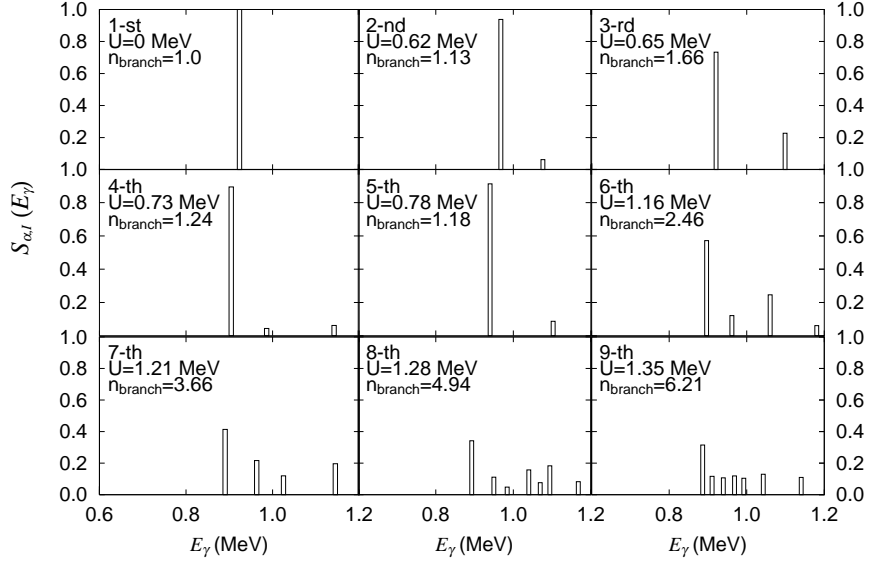}} \caption{The E2
strength distribution $S_{\alpha,I}(E_\gamma)$ from the lowest 9
states with $I^\pi=30^+$. The branch number $n_{\textrm {branch}}$
and excitation energy $U$ are put for each state.} \label{tran1}
\end{figure}

\begin{figure}
\epsfxsize=12.0cm \centerline{\epsffile{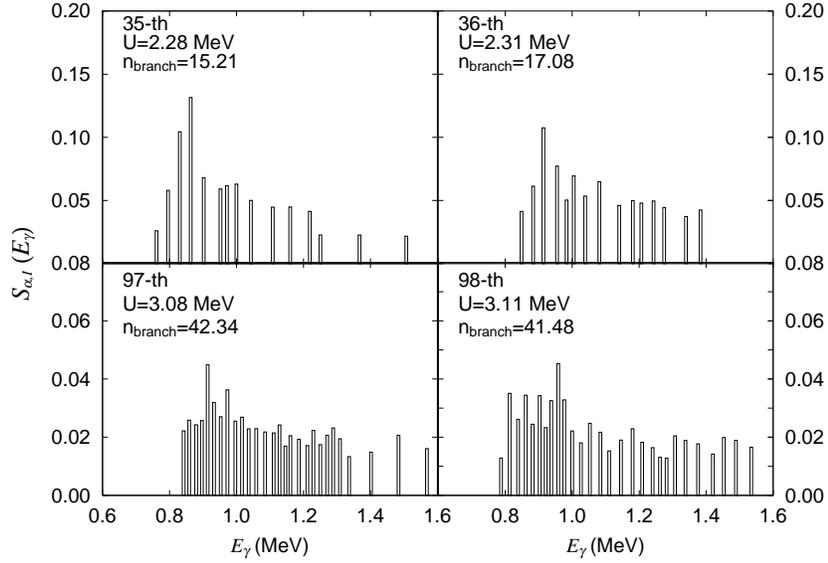}} \caption{The
strength distribution $S_{\alpha,I}(E_\gamma)$ for the stretched
E2 decays from the 35th and 36th, 97th and 98th excited states
with $I^\pi=30^+$.} \label{tran2}
\end{figure}

In order to indicate the E2 transition property more precisely,
the strength function $S_{\alpha,I}(E_\gamma)$ is shown in Fig.
\ref{tran1} for the lowest 9 states with $I^\pi=30^+$. The branch
number $n_{\textrm {branch}}$ and excitation energy $U$ are put
for each state. The E2 transition associated with the first $30^+$
state exhausts most of the total strength at gamma-ray energy 0.92
MeV. The $30_2^+$, $30_3^+$, $30_4^+$ and $30_5^+$ states show
essentially the same E2 distributions except for slight
difference in the gamma-ray energy. The E2 decays from the
$30_6^+$, $30_7^+$, $30_8^+$ and $30_9^+$ states display a
different E2 strength distribution, being fragmented over several
transitions, each of which carries a rather weak strength. Fig.
\ref{tran2} displays the quantity $S_{\alpha,I}(E_\gamma)$
associated with the states $30_{35}^+$ and $30_{36}^+$ lying at
excitation energy U $\sim$ 2.3 MeV, and the states $30_{97}^+$ and
$30_{98}^+$ lying at U $\sim$ 3.1 MeV. The E2 strength
distribution at U $\sim$ 2.3 MeV has about 16 branches, while the
number of branches becomes around 42 at U $\sim$ 3.1 MeV. It
indicates that the degree of the mixing between many-particle
many-hole configurations becomes stronger as excitation energy
increases, which is the basic mechanism to form the damping of
collective rotational motion.

\begin{figure}
\epsfxsize=8.0cm \centerline{\epsffile{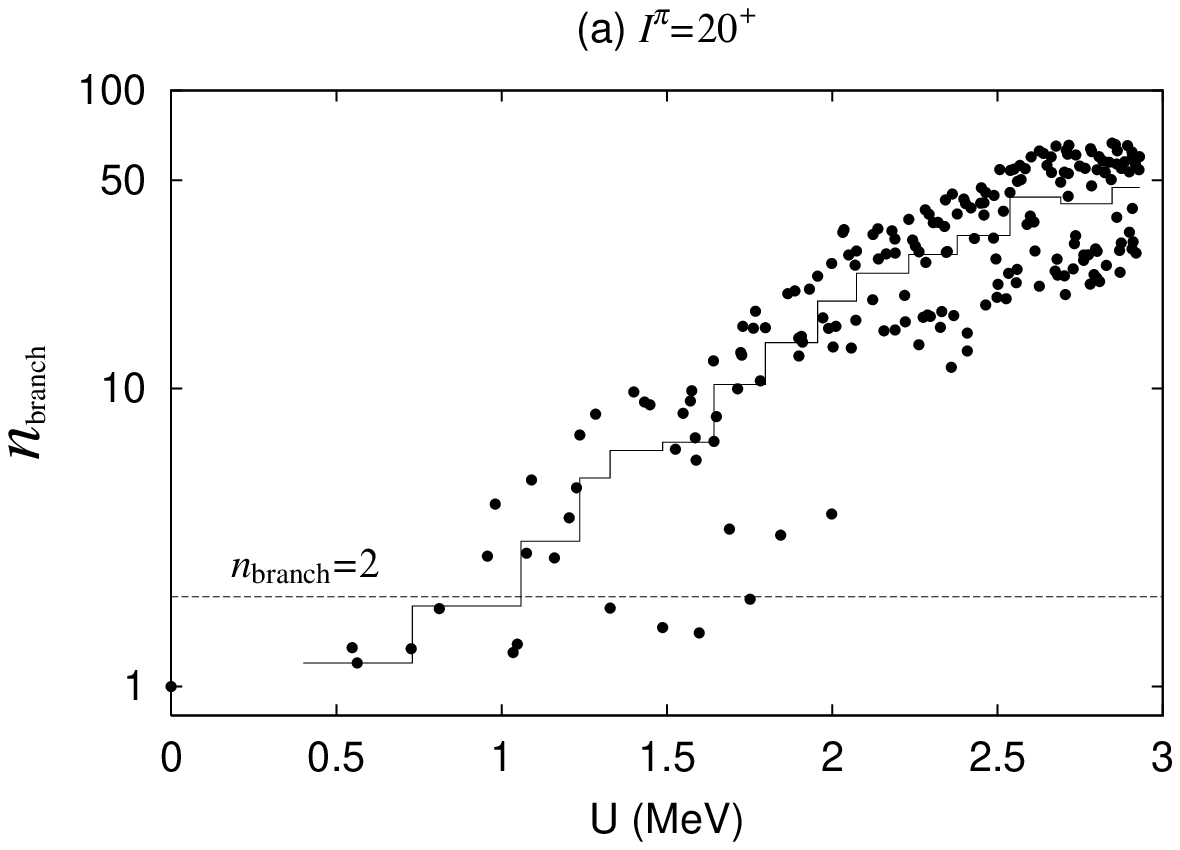}}
\centerline{\epsffile{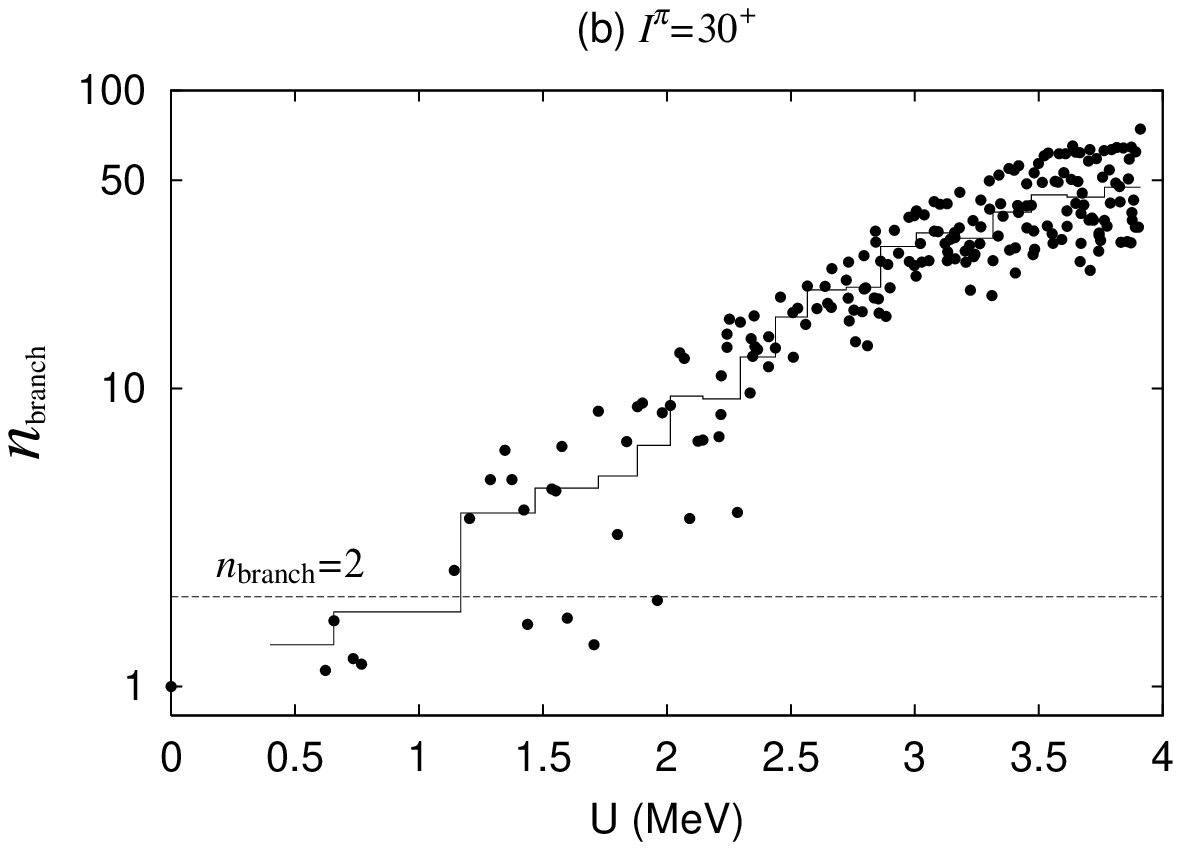}}
\centerline{\epsffile{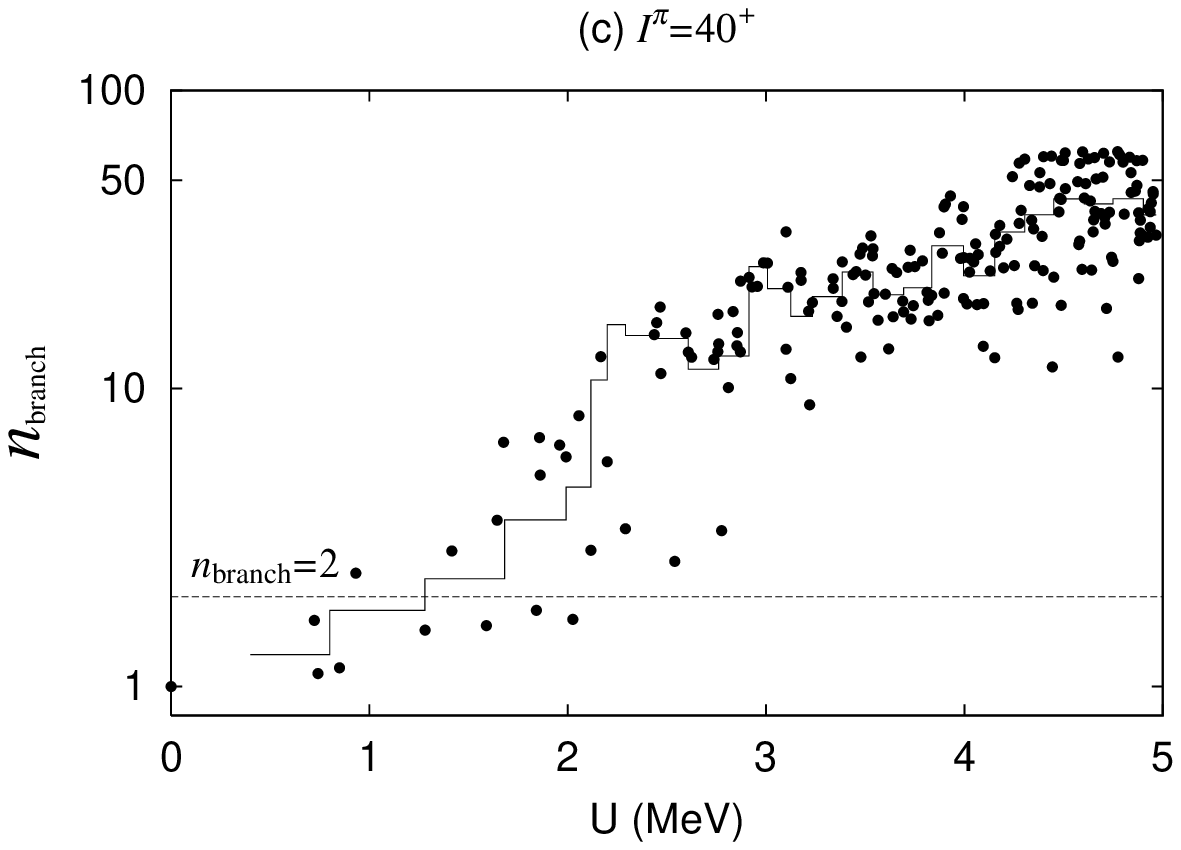}} \caption{The branching number
$n_{\textrm {branch}}$ for (a) $I^\pi=20^+$, (b) $I^\pi=30^+$ and
(c) $I^\pi=40^+$ as a function of excitation energy $U$. The
histogram gives the average branch number within the energy bins.
The horizontal line shows $n_{\textrm {branch}}$=2 used to define
the onset of rotational damping.} \label{branch}
\end{figure}

As the excitation energy increases, the rotational band structure
gradually disappears and rotational damping takes place. Branching
number is a key quantity to measure where rotational damping takes
place and the degree of configuration mixing. The dependence of
branching number on excitation energy is shown in Fig.
\ref{branch} for spins (a) $I^{\pi}=20^+$, (b) $I^{\pi}=30^+$ and
(c) $I^{\pi}=40^+$. The histogram gives the average branch number
within the energy bins. One may observe that branching number
increases with excitation energy. Using the criterion $n_{\textrm
{branch}}=2$ for the onset of rotational damping, the onset energy
is predicted to be around excitation energy $U \sim$ 1.1 MeV above
yrast line.

It should be noted, although the onset energy defined by condition
$n_{\textrm {branch}}=2$ tells approximately where the rotational
damping takes place, the transition from the region of rotational
bands to rotational damping is not very sharply at the onset
energy, but develops gradually as the excitation energy increases.
As shown in Fig. \ref{energy}, there exist some rotational bands
in the region of rotational damping, where the excitation energy
is higher than the onset energy. These band structures are
surrounded by states which do not have strong transitions. It
indicates that the rotational band structure partly remains even
in the region of rotational damping.  Such feature of rotational
bands is also displayed in Fig. \ref{branch}, where there exist
states whose branching number is smaller than 2 at higher
excitation energy.

\begin{figure}
\epsfxsize=8.0cm \centerline{\epsffile{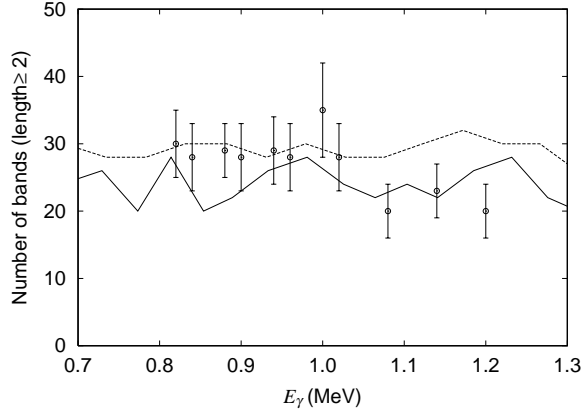}} \caption{ The
calculated number of bands together with the experimental
effective number of paths for nucleus $^{168}$Yb \cite{TD96} as a
function of the average transition gamma-ray energy. The solid
line is calculated with the criterion $S_{\alpha I+2,
\alpha^\prime I}>0.707$ while the dashed line represents the
result with condition $n_{\textrm {branch}}<2$.} \label{band1}
\end{figure}

Due to the onset of rotational damping, the number of rotational
bands existing in a given nucleus is thus limited and gives a
quantitative measure of rotational damping. Here, the number of
rotational bands corresponds to the experimental effective number
of paths which can be obtained from $E_{\gamma}\times E_{\gamma}$
spectrum by the fluctuation analysis method \cite{BH92,TD96}.
Theoretically the number of rotational bands is defined by the
number of states with strong E2 transition, which satisfies
condition $S_{\alpha I+2, \alpha^\prime I}>0.707$ or $n_{\textrm
{branch}}<2$ with at least two consecutive steps $I+2\rightarrow
I\rightarrow I-2$ of E2 decays~\cite{MM971}. Since the
configuration of valence nucleons occupancy in present model
corresponds to rare-earth nucleus $^{168}$Yb considered as a core
$^{160}$Yb plus eight valence neutrons and in our calculation the
moment of inertia is taken as the value of rare-earth nucleus, we
will make a qualitative comparison between our calculated number
of rotational bands and the experimental data for nucleus
$^{168}$Yb, though the model space in present calculation is not
really realistic for nucleus $^{168}$Yb. For this nucleus there
exists experimental data from the analysis of quasi-continuum
gamma-spectra as well as data from discrete spectra identifying
the rotational bands up to spin $I$ $\sim$ 40 \cite{AF95,JRB93}.
Fig. \ref{band1} shows the calculated number of rotational bands
as well as the experimental effective number of paths \cite{TD96}
for nucleus $^{168}$Yb. The solid line represents the result with
the criterion $S_{\alpha I+2, \alpha^\prime I}>0.707$ while the
dashed line is calculated with condition $n_{\textrm {branch}}<2$.
The horizontal axis denotes the average gamma-ray energy
$E_\gamma=(E_{\gamma _1}+E_{\gamma _2})/2$, where $E_{\gamma _1}$
is the transition energy for $I+2\rightarrow I$ and $E_{\gamma
_2}$ for transition $I\rightarrow I-2$. It is clear that the two
conditions give essentially the same number of rotational bands
around 30, and the theoretical calculation agrees well with the
experimental result in all the gamma-ray energy range.

\begin{figure}
\epsfxsize=8.0cm \centerline{\epsffile{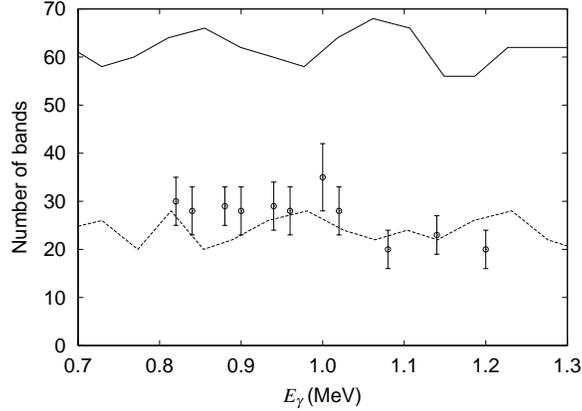}} \caption{The
calculated number of bands with and without pairing correlation
together with the experimental result as a function of the average
gamma-ray energy. The solid line is the result with pairing
strength $G=0.0$ MeV, while the dashed line is with the standard
pairing $G=0.45$ MeV. The criterion $S_{\alpha I+2, \alpha^\prime
I}>0.707$ has been used to obtain the number of rotational bands.}
\label{band2}
\end{figure}

In order to study the effect of pairing correlation between
valence nucleons on rotational damping, Fig. \ref{band2} displays
the calculated number of rotational bands with and without pairing
correlation, together with experimental result. The solid line
represents the result without pairing interaction between valence
nucleons, while the dashed line is with the standard pairing
$G=0.45$ MeV.  The criterion $S_{\alpha I+2, \alpha^\prime
I}>0.707$ has been used to obtain the number of rotational bands.
It is seen that the calculated number of bands becomes much larger
when the pairing is turned off. In other words, the rotational
band structure has strengthened and rotational damping has
weakened when pairing $G=0.0$ MeV. This indicates that pairing
correlation has a significant effect on the damping of rotational
motion.

\begin{figure}
\epsfxsize=8.0cm
\centerline{\epsffile{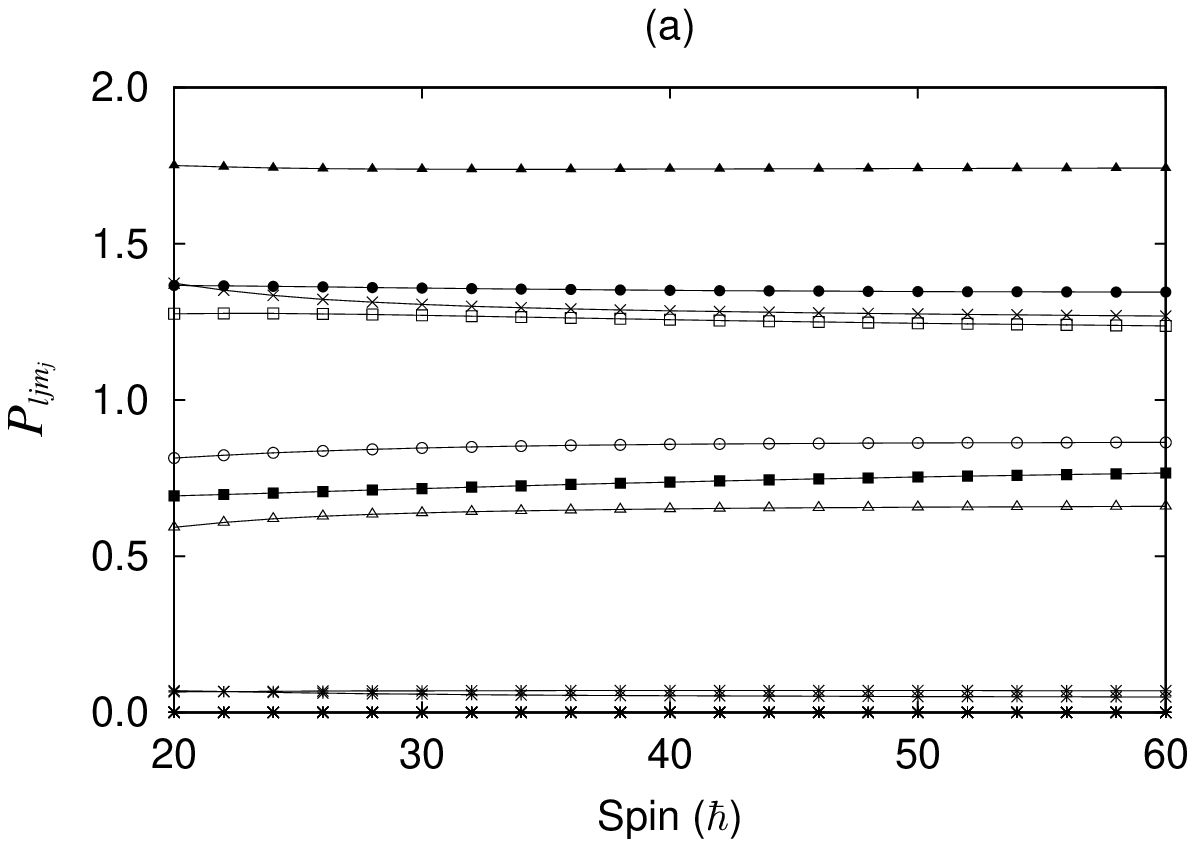}}
\centerline{\epsffile{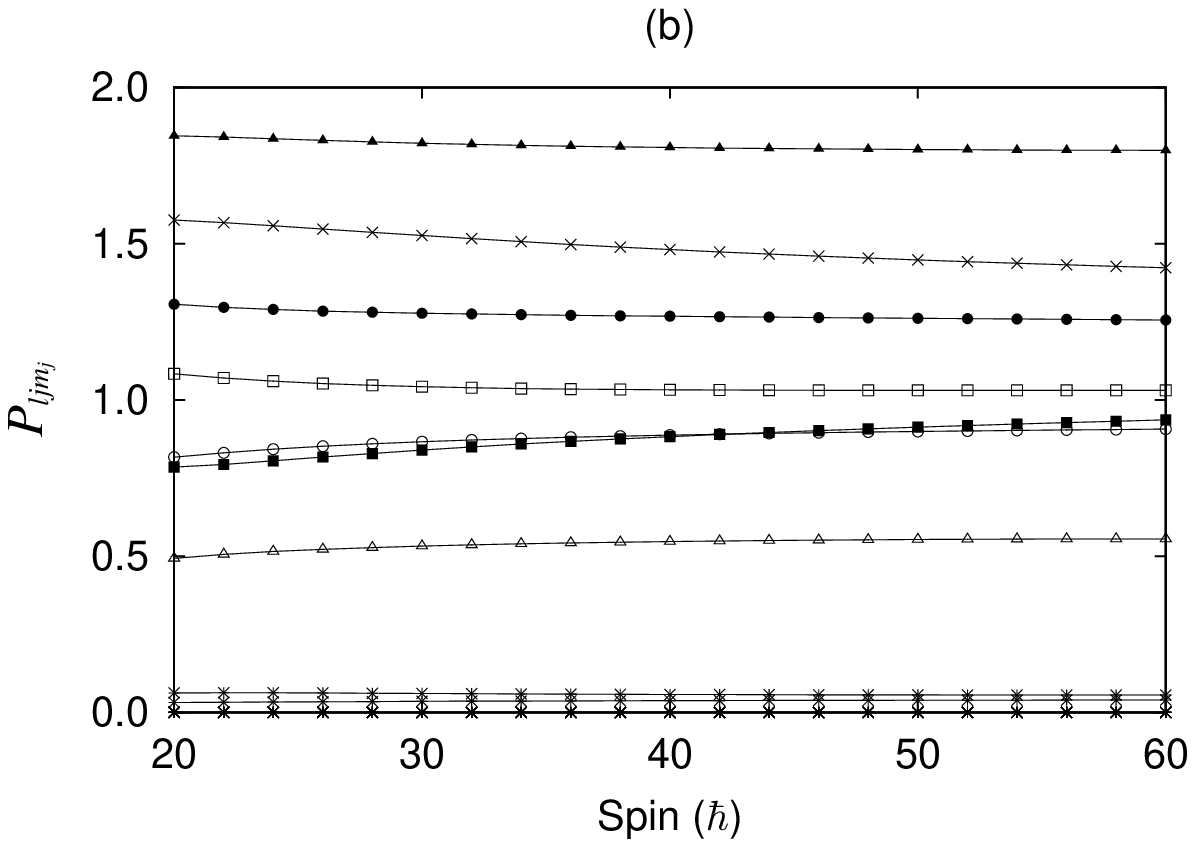}}
\centerline{\epsffile{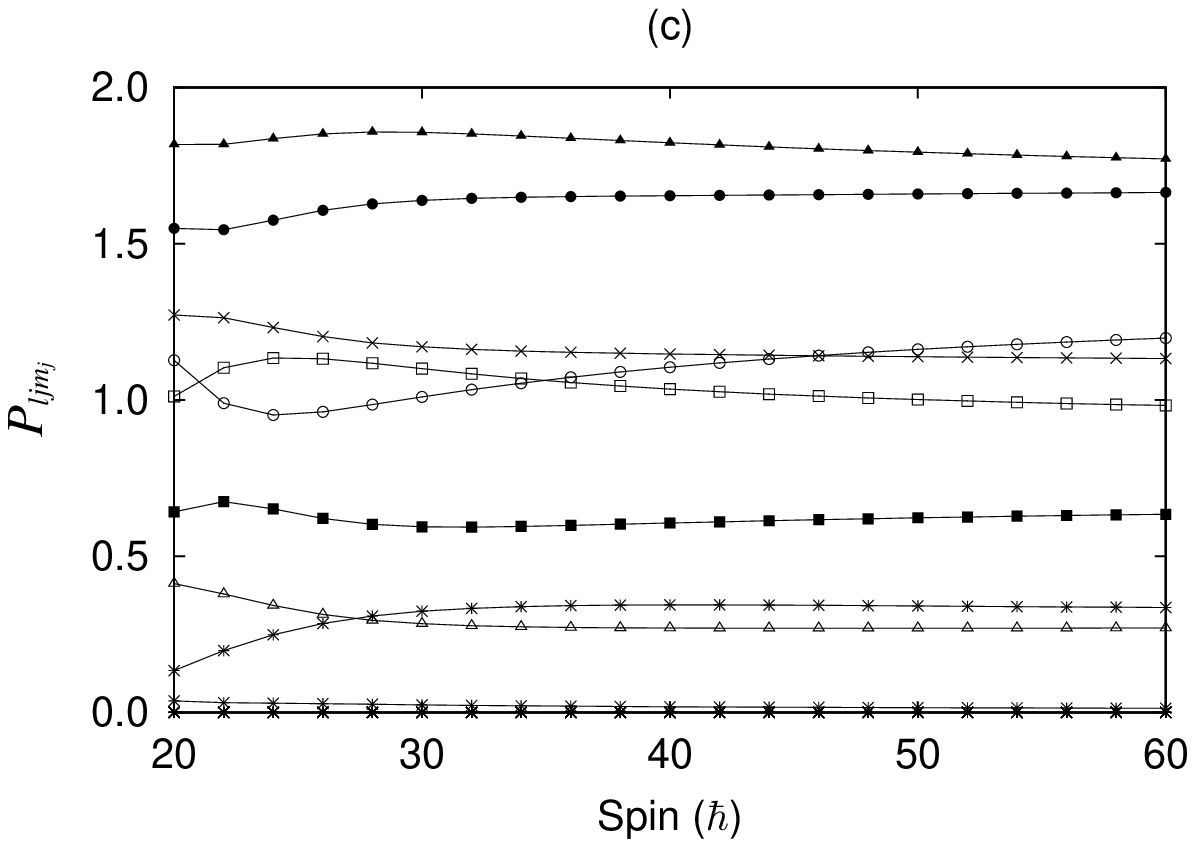}}
\caption{The same as in Fig.~\ref{Occpro1}, but without pairing correlation.}
\label{Occpro2}
\end{figure}

To understand the nature of nuclear state without pairing
correlation, Fig.~\ref{Occpro2} gives the occupation number of
particle of the same bands as in Fig.~\ref{Occpro1}, but with
pairing strength $G=0.0$ MeV. One may observe for these bands the
dependence of occupancy on spin becomes weaker in the case of
without pairing correlation. Especially in low angular momentum,
there is much difference between the occupancy with and without
pairing. These results are consistent with the calculation in
Fig.~\ref{band2}, where the number of rotational bands becomes
larger when paring $G=0.0$ MeV. It indicates that pairing
correlation has an important effect on the nature of nuclear
states especially in low angular momentum.

It should be mentioned that pairing correlation between valence
nucleons has contributions to both diagonal and off-diagonal
Hamiltonians. Diagonal Hamiltonian characterizing the property of
mean-field favors the rotational band structure and retards the
rotational damping, whereas the off-diagonal Hamiltonian
characterizing the quantum fluctuations coming from residual
interaction causes the mixing of many-particle many-hole
configurations and prefers the rotational damping. Therefore, the
delicate balance between the competition of diagonal and
off-diagonal components of pairing correlation has determined the
appearance of rotational damping. It is concluded that the
off-diagonal components of pairing correlation (two-body residual
interaction) play an important role in the appearance of
rotational damping.

\section{conclusions}
The damping of collective rotational motion is discussed in a
multi-$j$ shell particles-rotor model, in which the angular
momentum is strictly conserved and the pairing correlation is
included explicitly. It is found that the rotational damping takes
place at about 1.1 MeV above yrast line, and the number of states
which form rotational band structure is thus limited. The onset
energy in present calculation is similar as the theoretical
prediction $U \sim 0.8$ MeV in cranked Nilsson mean-field combined
two-body residual interaction. The calculated number of rotational
bands around 30 is in a qualitative agreement with experimental
result in the gamma-ray energy range. The onset of rotational
damping takes place quite gradually as a function of excitation
energy. Even in the region of rotational damping, there still
remains part of discrete rotational band structure. Our calculation clearly
indicates that the pairing correlation has an important effect on
the nature of nuclear states, especially in low angular momentum.
It is found that the calculated number of bands becomes much larger in the
case of pairing strength $G=0.0$ MeV. 
The pairing correlation between valence particles has significant
effect on the appearance of rotational damping. It is noted that the
pairing correlation and the exact treatment on angular momentum
coupling in PRM play an important role to provide a description on
rotational damping. However in the cranked Nilsson mean-field
combined two-body residual interaction, the rotational damping is 
attributed to the high-multipole component of two-body residual 
interaction~\cite{MM971}. Considering that the angular
momentum in Ref.~\cite{MM971} is not treated properly and the valence
nucleons states are schematic in present model, more works along
these lines are necessary to understand better the mechanism of
rotational damping.

\begin{acknowledgments}
This work was supported in part by the Japan Society for the
Promotion of Science (JSPS) and the China National Natural Science
Foundation (CNSF) as the bilateral program between Japan and
China. The partial supports from the Major State Basic Research
Development Program Under Contract Number G2000077407, the
National Natural Science Foundation of China under Grant No.
10025522, 10221003, 10375001, and 10435010, the Doctoral Program
Foundation from the Ministry of Education in China, and the
Knowledge Innovation Project of the Chinese Academy of Sciences
under Grant No. KJCX2-SW-N02 are acknowledged.
\end{acknowledgments}

\newpage
\begin{table}
\caption{\label{tab1} The relative energy $R_j$ between shell $h_{9/2}$ and other
shells $i_{13/2}$, $p_{3/2}$, $f_{5/2}$ and $p_{1/2}$ in unit of $\hbar\omega_0$.
The level scheme is shown on the left in Fig.~\ref{ref1}.}
\begin{ruledtabular}
\begin{tabular}{c|cccc}
$R_j$   &  $i_{13/2}$  &  $p_{3/2}$  &  $f_{5/2}$ &  $p_{1/2}$ \\
\hline
$h_{9/2}$ &  0.1 & 0.3 & 0.35 & 0.49
\end{tabular}
\end{ruledtabular}
\end{table}

\end{document}